\documentclass[pdftex]{article}
\usepackage{amsmath}
\usepackage{amsthm}
\usepackage{amssymb}
\usepackage{graphicx}
\usepackage{braket}
\usepackage{verbatim}
\usepackage[margin=1in]{geometry}
\usepackage{leftidx}
\usepackage{wrapfig}
\usepackage{multicol}
\usepackage{array}
\usepackage{subfig}
\usepackage[T1]{fontenc}
\usepackage{bbm}

\title{Tunneling in cosine potential with periodic boundary conditions.}
\author{Zbigniew Ambrozi\'nski}

\begin{document}
\maketitle

\begin{abstract}
In this paper we discuss three methods to calculate energy splitting in cosine potential on a circle, Bloch waves, semi--classical approximation and restricted basis approach. While the Bloch wave method gives only a qualitative result, with the WKB method we are able to determine its unknown coefficients. The numerical approach is most exact and enables us to extract further corrections to previous results.
\end{abstract}

\section{Introduction}
In this article we study tunneling in quantum mechanical cosine potential with periodic boundary conditions. Hamiltonian of this system is given by the formula
\begin{align}
H=\frac{1}{2}P^2+\frac{1}{4\pi^2g}\left(1-\cos\left(2\pi\sqrt g X\right)\right)
\end{align}
with $X\in(0,g^{-1/2}K)$ where $K$ is the number of minima and with periodic boundary conditions. For shorter notation we put $V(x)=\frac{1}{4\pi^2}\left(1-\cos(2\pi x)\right)$. Within perturbative calculus vacuum energy of the system is degenerate. Each ground state is localized in different minimum. Tunneling is responsible for splitting of the energies. In general, this effect cannot be studied analytically. One can give a first approximation to the splitting using semiclassical--approximation (or WKB approximation) which was developed by G. Wentzel, H. Kramers and L. Brillouin in 1926. In this approach, one finds that the ground energy is shifted by a quantity which is a nonperturbative function of coupling constant. This classical solution is called an instanton.

Validity of the instanton calculus is limited to systems with widely separated minima with a large potential barrier in between. Nevertheless, it has a vast range of applications to modern field theories. In Yang--Mills theory with $SU(2)$ symmetry group zero energy states are pure gauge (see \cite{ITEP,Schafer}). Such fields can be viewed as mappings of $SU(2)$ group into itself and they can be divided into sectors due to topological properties. These sectors are labeled by topologically invariant Pontryagin index which takes only integer values. Clearly, there is no continuous pure gauge transformation connecting two topological vacua belonging to different sectors. There are however configurations with non vanishing field strength which interpolate between different topological vacua $\ket{n}$, where $n$ is the Pontryagin index. A special configuration is the BPST instanton which is the minimal action path connecting $\ket{n}$ and $\ket{n+1}$ in Euclidean space. Presence of the BPST instanton has dramatic consequences for structure of the real vacuum. The true vacuum of the theory is not a single topological vacuum $\ket{n}$ but a superposition of all such vacua $\ket{\theta}=\sum_n e^{in\theta}\ket{n}$, where $\theta$ is called vacuum angle.

Such system can be modeled in one dimensional quantum mechanics by a periodic potential. According to Bloch theorem, the energy spectrum consists of continuous bands. Each eigenenergy is labeled by an angle $\theta$ and energy states are superpositions of states localized in single minima. Unlike in the quantum mechanical case, in Yang--Mills theory only one value of $\theta$ is admissible due to superselection rule. No energy bands are present and there is a mass gap between the vacuum and the first excited state. The vacuum angle in QCD is responsible for violating CP symmetry. On the other hand, there is no experimental evidence for CP breaking which imposes a limit on the angle $|\theta|<10^{-9}$ \cite{Crewther}.

The periodic potential in the weak coupling limit shares many features of the double well potential which was extensively studied since 1960s \cite{Bender,Caswell,Meurice,ZJ,ZJ1,Bogomolny}. It was discovered that there are further corrections to the WKB approximation which can be derived from modified Bohr--Sommerfeld quantization condition \cite{ZJ2}. They come from from multi--instanton molecules (i.e. a classical path in Euclidean space which is composed of instantons that are close to each other) and contribute to the ground energy much weaker than a single instanton. Secondly, each instanton molecule contribution (including single instantons) is multiplied by a series, which is presumably asymptotic. Moreover, as stated recently by M. \"Unsal \cite{Unsal}, interactions between instantons can heal non Borel summability of perturbation series for potentials with degenerate global minima. All above mentioned effects take place also case of periodic potential.

One has to keep in mind that instanton considerations neglect perturbative contributions to energies, which are much larger. Secondly, there are higher order corrections to instanton contributions which become significant at stronger coupling. It is understandable that there is a need to verify statements concerning instantons and see in what regime of the coupling constant the instanton picture is valid. In quantum mechanics there is a very efficient approach, called restricted basis method \cite{Wosiek:2002}. It originates from the variational Tamm--Dancoff method \cite{Dancoff}. In the restricted basis approach one takes basis states $\ket{n}$ with $n$ smaller than a certain large cut--off and finds energies in this subspace. A price for taking many states, rather than a few trial functions as in original Dancoff paper, is that the calculations have to be performed numerically. On the other hand, it is very efficient in one dimensional quantum mechanics. Indeed, a good convergence of energies with growing cut--off has been observed numerically \cite{Wosiek}. Accuracy of this technique is limited only by precision of computations and the size of the space. Apart from these limitations, this method is exact and provides then a powerful tool for testing WKB approximation.

Plan of this paper is the following. In Section \ref{sec:Bloch} we present the traditional approach to periodic potential with the use of Bloch waves. In Sections \ref{sec:WKB} we apply the instanton calculus in order to find low energies of the system. Results of the semi--classical approximation are consistent with preceding section and give more quantitative answer. Section \ref{sec:Fock} provides the most complete answer. It confirms results obtained by the WKB method and show further perturbative corrections. A few coefficients of the series are extracted from numerical data.

\section{Bloch waves}\label{sec:Bloch}
Let us first consider system of infinite size, i.e. $K=\infty$. According to Bloch theorem, the lowest energies of the system form a continuous band of width $\Delta$. They are usually parameterized by an angle:
\begin{align}\label{eq:Bloch_energies}
E(\theta)&=\bar E-\frac{\Delta}{2}\cos\theta,&\theta\in(-\pi,\pi).
\end{align}
Corresponding wavefunction is a plane wave modulated by a periodic function $u_\theta(x)$ with period $g^{-1/2}$, which is the period of the potential:
\begin{align}\label{eq:Bloch_waves}
\psi_\theta(x)=\exp\left(i\theta xg^{1/2}\right)u_\theta(x).
\end{align}
Both, $\bar E$ and $\Delta$ depend on specific shape of the potential and are not determined by the Bloch theorem.

For finite $K$ the wavefunction satisfies periodic boundary condition $\psi_\theta(0)=\psi_\theta(Kg^{-1/2})$ and only discrete values of $\theta$ are allowed.
They are: $\theta=2\pi j/K$ with $j=-K/2+1,\ldots,K/2$ for even $K$ and $j=-(K-1)/2,\ldots,(K-1)/2$ for odd $K$. One can see that energies are doubly degenerate: $E(\theta)=E(-\theta)$ for all $\theta$ except $\theta=0,\pi$.

Bloch theorem gives a qualitative answer to what are low energies of the system. Still, values of the mean energy of the band $\bar E$ and width $\Delta$ need to be found.
\section{WKB approximation}\label{sec:WKB}
In this section we will show how lowest energies of the hamiltonian may be obtained in the well known instanton calculus. Analogous calculation was given in detail for the double well potential by S. Coleman in \cite{Coleman}. A general discussion concerning periodic potentials can be found in \cite{Coleman2}. These methods will be applied to the special case of cosine potential with periodic boundary conditions. The transition amplitude from minimum $x=0$ to $x=g^{-1/2}k,\ k=0,\ldots,K-1$ in Euclidean time $T$ may be expressed as a path integral
\begin{align}\label{eq:WKB_integral}
\braket{g^{-1/2}k|e^{-TH}|0}=\mathcal N\int \mathcal D[x(\tau)]e^{-S_E[x(\tau)]},
\end{align}
where $\mathcal N$ is a normalization factor. The integral is over trajectories which satisfy boundary conditions $x(-T/2)=0,\ x(T/2)=g^{-1/2}k$. The Euclidean action $S_E$ is
\begin{align}
S_E[x(\tau)]&=\int_{-T/2}^{T/2}d\tau\left(\frac{1}{2}\dot x(\tau)^2+g^{-1}V(g^{1/2}x(\tau))\right).
\end{align}
A trajectory $\bar x(\tau)$ that minimizes the action for $k=1$ is called an instanton. For $T\to\infty$ analytical solution yields $\bar x(\tau)=\frac{2}{\pi}g^{-1/2}\arctan(e^\tau)$ with $S_0\equiv S_E[\bar x(\tau)]=2/\pi^2g$. For $k>1$ minimal solutions are composed of many such instantons and are called multi--instanton paths. Each instanton of such path connects two neighbor minima and begins where its predecessor ended. The integral (\ref{eq:WKB_integral}) is calculated in gaussian approximation around multi--instanton paths and yields
\begin{align}
\mathcal N\int \mathcal D[x(\tau)]e^{-S_E[x(\tau)]}=\frac{1}{\sqrt\pi}e^{-T/2}\sum_{n=0}^\infty c_{n,k}\frac{1}{n!}\left(\sqrt\frac{S_0}{2\pi}T\right)^n\mathcal K^ne^{-nS_0}.
\end{align}
The coefficient $c_{n,k}$ is the number of topologically different $n$--instatnon configurations which satisfy appropriate boundary conditions. Constant $\mathcal K$ is
\begin{align}
\mathcal K&= \left(\frac{\det\left[-\frac{d^2}{d\tau^2}+1\right]}{\det{}'\left[-\frac{d^2}{d\tau^2}+V''(g^{-1/2}\bar x(\tau))\right]}\right)^{1/2}=2.
\end{align}
Determinants in the above formula are understood as products of all eigenvalues and symbol $'$ indicates that the lowest eigenvalue is omitted. Calculation of  $\mathcal K$ is postponed to Appendix \ref{sec:determinants}.

Let us now calculate $c_{n,k}$. Obviously, $c_{0,k}=\delta_{0,k}$ and the boundary condition is $c_{n,K}\equiv c_{n,0}$. Let $x(\tau)$ be an n--instanton path ending at $k$th minimum. Then the penultimate instanton has to end at minimum $k-1$ or $k+1$. Therefore, $c_{n,k}=c_{n-1,k-1}+c_{n-1,k+1}$. Recursive relations can be written as a matrix equation $c_{n,k}=\sum_lR_{kl}c_{n-1,l}$ with $R_{kl}=\delta_{k-1,l}+\delta_{k+1,l}$. Let us notice that eigenvalues of $R$ are $\lambda_j=2\cos(2\pi j/K)$ with corresponding eigenvector $(v_j)_k=\exp(2\pi i jk/K)$. Then, $c_{n,k}=\sum_j \alpha_j(v_j)_k\lambda_j^n$. Coefficients $\alpha_j$ are determined from initial condition $c_{0,k}=\delta_{0,k}$ and yield $\alpha_j=1/K$ for all $j$.
Finally,
\begin{align}\label{eq:WKB_amplitude}\begin{split}
\braket{g^{-1/2}k|e^{-TH}|0}&=
\frac{1}{K\sqrt\pi}e^{-T/2}\sum_{n=0}^\infty\sum_{j=0}^{K-1} e^{2\pi i jk/K} \frac{1}{n!}\left(\cos(2\pi j/K)\mathcal K\sqrt\frac{2S_0}{\pi}e^{-S_0}T\right)^n\\
&=\frac{1}{K\sqrt\pi}e^{-T/2}\sum_{j=0}^{K-1} e^{2\pi i jk/K} \exp\left(4\cos(2\pi j/K)\frac{1}{\sqrt{\pi^3g}}e^{-2/\pi^2g}T\right).
\end{split}\end{align}

One can use the identity $\ket{E}\bra{E}=\mathbbm 1$ to expand the amplitude $\braket{g^{-1/2}k|e^{-TH}|0}$ as follows:
\begin{align}
\braket{g^{-1/2}k|e^{-TH}|0}=\sum_E \braket{g^{-1/2}k|E}\braket{E|0}e^{-TE}.
\end{align}
By comparison with formula (\ref{eq:WKB_amplitude}) we extract energies and values of wavefunctions at minima
\begin{align}
E_j&=\frac{1}{2}-4\cos(2\pi j/K)\frac{1}{\sqrt{\pi^3g}}e^{-2/\pi^2g},\\
\braket{g^{-1/2}k|E_j}&=\frac{1}{\sqrt{K\sqrt\pi}}e^{2\pi i jk/K}.\label{eq:equation_for_amplitudes}
\end{align}
Note that $E_{j}=E_{K-j}$. Therefore, each energy for $j=1,\ldots,\lfloor\frac{K-1}{2}\rfloor$ is degenerate. The lowest energy $E_0$ is always non-degenerate and $E_{K/2}$ is non-degenerate for even $K$.

For infinite $K$ the parameter $\theta=\frac{2\pi j}{K}$ becomes continuous. Energies and eigenstates take the same form as in (\ref{eq:Bloch_energies}) and (\ref{eq:Bloch_waves}):
\begin{align}
E_\theta&=\frac{1}{2}-4\cos(\theta)\frac{1}{\sqrt{\pi^3g}}e^{-2/\pi^2g},\label{eq:WKB_energy}\\
\braket{g^{-1/2}k|\theta}&=\pi^{-1/4}e^{i\theta k}\label{eq:WKB_state}.
\end{align}
To each energy $E_\theta=E_{2\pi-\theta}$ correspond two states $\ket{\theta}$ and $\ket{2\pi-\theta}$.

According to [ZJ] there are perturbative corrections to formula (\ref{eq:WKB_energy}):
\begin{align}\label{eq:ZJ_expansion}
E_j&=\sum_{k=0}^\infty{a_kg^k}-2\cos(2\pi j/K)\sqrt\frac{2S_0}{\pi}e^{-S_0}\sum_{k=0}^\infty{b_kg^k}+\mathcal O(e^{-2S_0}\log(g))
\end{align}
with $a_0=\frac{1}{2},\ b_0=1$.

\section{Restricted basis approach}\label{sec:Fock}
Yet another technique, restricted basis method can be used to obtain eigenenergies of the Hamiltonian. It is performed after \cite{Wosiek}. Instead of Fock basis we use plane waves which are more convenient in case of periodic potential:
\begin{align}
\braket{x|n}&=\frac{g^{1/4}}{\sqrt K}\exp(2\pi i ng^{1/2}x/K).
\end{align}
The Hamiltonian is symmetric under $\mathbb Z_K$ group transformation. Let $T$ be the shift operator: $T\ket{x}=\ket{x+g^{-1/2}}$. Its action on the basis vectors is $T\ket{n}=\exp(-2\pi i n/K)\ket{n}$. Since $[H,T]=0$, the Hamiltonian $H$ can be diagonalized on each eigensubspace of $T$ separately. There are $K$ eigensubspaces of $T$, $\mathcal H_{j,K}=span\{\ket{n}_j,n\in\mathbb Z\}$ where $\ket{n}_j=\ket{j+nK}$. Then the Hamiltonian is an infinite tridiagonal matrix
\begin{align}
{}_j\!\braket{m|H|n}_j&=g\left(\frac{2\pi (j+nK)}{K}\right)^2\delta_{m,n}+\frac{1}{8\pi^2g}(2\delta_{m,n}-\delta_{m,n-1}-\delta_{m,n+1}),&m,n\in\mathbb Z.
\end{align}
Let us note that the hamiltonian $H$ in sector $\mathcal H_{j,K}$ is the same as in the sector $\mathcal H_{lj,lK}$ for any $l\in\mathbb N$. Therefore, the spectrum for $K=\infty$ contains all energies from sectors $\mathcal H_{j,K}$ with $p$ and $K$ being coprime integers. The set of lowest energies from all sectors form a dense set in an interval which is the continuous energy band for infinite $K$.

In order to obtain energies one has to introduce a cutoff $|n|<N$ and use numerical methods to find energies. Because the matrix is sparse, Arnoldi algorithm is very efficient.

\begin{figure}
\centering
\includegraphics[width=.8\textwidth]{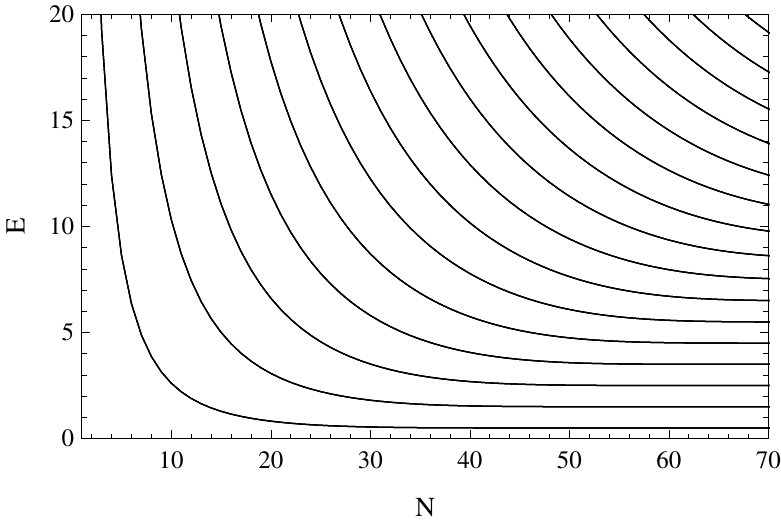}
\caption{Convergence of energies with growing cutoff $N$ for $K=4$ and $g=10^{-4}$. Each line represents four energies which are almost degenerate. For larger $g$ convergence is even faster.}
\label{fig:convergencegraph}
\end{figure}

One of the most important issues is rate of convergence. We shall note that in all known cases convergence of energies is exponential if the spectrum is discrete and it converges roughly like $N^{-1}$ if it is continuous. For each finite $K$ spectrum is discrete. For $K=\infty$ the spectrum is continuous. However, it is discrete in each sector $\mathcal H_{j,K}$ separately while there is an infinite number of sectors. Convergence in each sector is exponential. Energies as functions of cutoff $N$ are presented in Fig. \ref{fig:convergencegraph}.

We are interested in calculating width of the energy band $\Delta=E_{K/2}-E_0$, where $E_j$ is the ground energy in the sector $\mathcal H_{j,K}$. The smallest value of coupling constant used in our computations was $g=9.13\times 10^{-6}$. Precision of computations is determined by value of $\Delta$ for different $g$. For $g=9.13\times 10^{-6}$ it is $\Delta=2.6\times 10^{-9637}$. Needed cutoff was $N=13000$.

\section{Comparison of results}\label{sec:comparison}
We will first check the agreement between the Bloch wave approach and WKB method. By comparing (\ref{eq:Bloch_energies}) and (\ref{eq:WKB_energy}) one can read mean energy and width of energy band in (\ref{eq:Bloch_energies}):
\begin{align}
\bar E&=\frac{1}{2},\\
\Delta&=\Delta_{WKB}\equiv 8\frac{1}{\sqrt{\pi^3g}}e^{-2/\pi^2g}.
\end{align}
From (\ref{eq:WKB_state}) we can see that $\ket{\theta}$ is an eigenstate of the translation operator $T$ with eigenvalue $e^{i\theta}$. Thus, the wavefunction $\braket{x|\theta}$ has the same form as the Bloch wave (\ref{eq:Bloch_waves}).

\begin{figure}
\includegraphics[width=.8\textwidth]{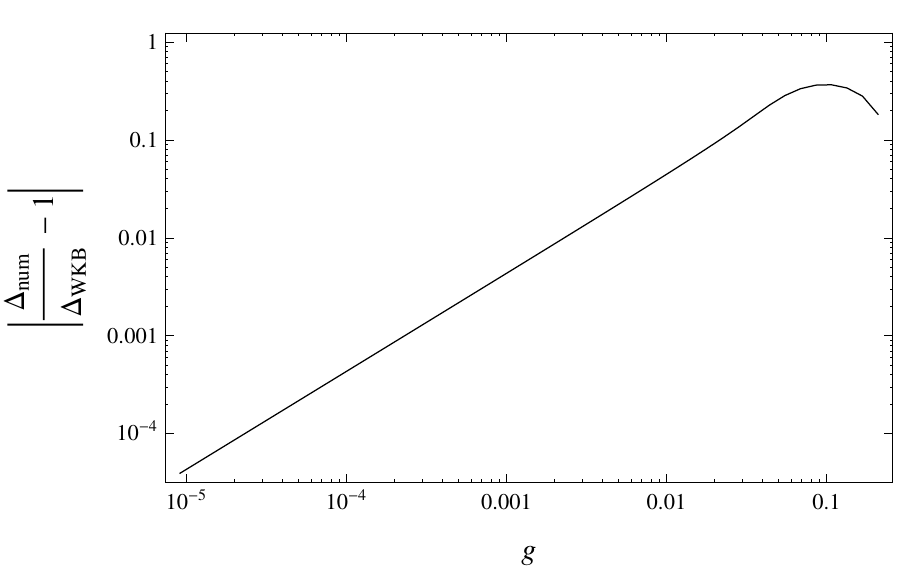}
\caption{Relative difference between semiclassical and numerical results for splitting of energies tends to $0$ as $g\to0$. Linear convergence on the log--log plot indicate that there are power corrections to the WKB prescription.}
\label{fig:relative_difference_graph}
\end{figure}

Comparison of the WKB and restricted basis methods requires more refined analysis. The semiclassical method neglects perturbative corrections to energies, which are seen in the other approach. Therefore, we will compare only width of the low energy band $\Delta$, which is purely nonperturbative. The semiclassical approximation is valid for small $g$ so we expect that $\Delta_{num}/\Delta_{WKB}-1\to 0$ as $g\to0$. This convergence is shown in Fig. \ref{fig:relative_difference_graph}.
We can read from (\ref{eq:ZJ_expansion}) that the ratio $\Delta_{num}/\Delta_{WKB}$ is a power series:
\begin{align}
\Delta_{num}/\Delta_{WKB}=\sum_{k=0}^\infty b_k g^k.
\end{align}
It turns out that numerical results are precise enough to extract several coefficients $b_k$. It appears that $b_k 2^{5k}\pi^{-2k}$ are integers well within error estimates. Their numerical values are

\begin{align}
\begin{array}{lr@{\ \pm\ }l}
k&\multicolumn{2}{c}{ 2^{5k}\pi^{-2k}b_k}\\
0&1&6.3\times 10^{-46}\\
1&-14&8.6\times 10^{-40}\\
2&-118&5.1\times 10^{-34}\\
3&-3588&1.8\times 10^{-28}\\
4&-150010&4.0\times 10^{-23}\\
5&-7665092&6.2\times 10^{-18}\\
6&-454322300&6.7\times 10^{-13}\\
7&-30378374408&5.3\times 10^{-8}\\
8&-2 253 225 850 810&3.0\times 10^{-3}\\
9&-183  329 494 073 630&1.2\times 10^{2}
\end{array}\label{eq:coefficients}\end{align}

Perturbative corrections can be extracted from any $E_j$. However, perturbative corrections can be found more easily up to higher orders by a Rayleigh-Schr\"odinger perturbation theory in a selected minimum of the potential.

\begin{figure}
\centering
\includegraphics[width=.8\textwidth]{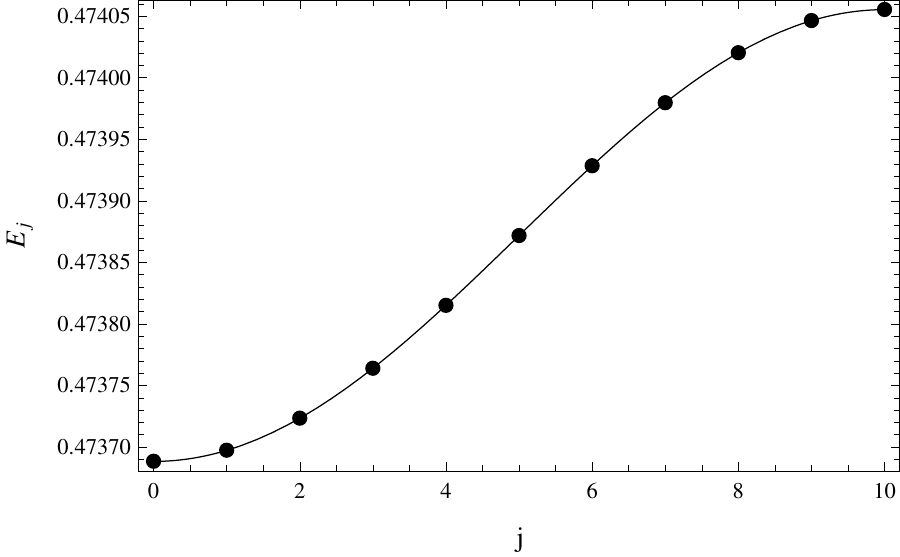}
\caption{Dependence of energies on the index $j$ for $K=20$, $g=0.02$. There are 2 nondegenerate and 9 doubly degenerate energies. Dots represent numerical data. The continuous line is function $E(\theta)=\bar E-\frac{\Delta}{2}\cos\theta$ with $\theta=2\pi j/K,\ \bar E=E_{K/4},\ \Delta=E_{K/2}-E_0$. Agreement is weaker for larger values of $g$.}
\label{fig:mean_energies}
\end{figure}

We also check whether energy dependence on the parameter $\theta=2\pi j/K$ agrees with (\ref{eq:Bloch_energies}). Results are presented in Fig. \ref{fig:mean_energies}. One can see a very nice agreement for $g=0.02$. It is violated for larger values of coupling constant.

Structure of wavefunctions is consistent with Bloch waves. Indeed, since the state $\ket{E_j}$ is in the sector $\mathcal H_{j,K}$, it is an eigenvector of translation operator $T$ corresponding to eigenvalue $\exp(-2\pi i j/K)$. Therefore the wavefunction $\psi_j(x)=\braket{x|E_j}$ satisfies
\begin{align}
\braket{x+g^{-1/2}|E_j}=\braket{x|T^\dagger|E_j}=\exp(2\pi i j/K)\braket{x|E_j},
\end{align}
i.e. it is a periodic function modulated by a plane wave.

\section{Summary}
We have considered three approaches to find low energies for the cosine potential in the small coupling limit -- Bloch waves, instanton calculus and restricted basis approach. They all gave consistent results, both for finite and infinite $K$. Wavefunctions corresponding to each energy also have the same form. While the Bloch theorem gave us only qualitative results, we found values of the mean energy $\bar E$ and band width $\Delta$ with the use of instanton calculus.

Next, we established agreement of the WKB approximation and the restricted basis space method. It turned out that values of $\bar E$ and $\Delta$ determined by the former method are only zero order approximation. We found perturbative corrections to $\Delta$ with the numerical technique while calculating corrections to $\bar E$ is trivial.

Another interesting issue is summability of aforementioned series. The perturbative series of ground energy is known to be asymptotic and thus not summable. Though, its Borel sum can be given a meaning when one includes correction due to interactions of instantons. The author addresses this problem in \cite{tobepublished}.

The series $\sum b_k g^k$ also appears to be asymptotic. From coefficients (\ref{eq:coefficients}) one can estimate the asymptotic behavior $b_k\approx-1.1\times2.8^kk!$. Borel sum of this series may be given a meaning after including corrections due to interaction of triples of instantons. General scheme of resummation for all orders is proposed in \cite{Unsal}.

\subsection*{Acknowledgements}

This work was supported by Foundation for Polish Science MPD Programme co-financed by the European Regional Development Fund, agreement no. MPD/2009/6.

\appendix
\section{Calculation of the determinant}\label{sec:determinants}
We will now show the method of calculating ratio of determinants
\begin{align}
\mathcal K&= \left(\frac{\det\left[-\frac{d^2}{d\tau^2}+1\right]}{\det{}'\left[-\frac{d^2}{d\tau^2}+V''(g^{-1/2}\bar x(\tau))\right]}\right)^{1/2}.
\end{align}
Let $L_i=-\frac{d^2}{d\tau^2}+W_i(\tau)$ and let $\psi^{(i)}_\lambda(\tau)$ satisfy
\begin{align}\label{eq:Liouville_eq}
L_i\psi^{(i)}_\lambda(\tau)&=\lambda\psi^{(i)}_\lambda(\tau),
&\psi^{(i)}_\lambda(-T/2)=0,\ \left.\frac{d}{d\tau}\psi^{(i)}_\lambda(\tau)\right|_{\tau=-T/2}=1.
\end{align}
We say that $\lambda$ is an eigenvalue of $L_i$ if $\psi^{(i)}_\lambda(T/2)=0$. It was shown in \cite{Coleman} that for two bounded functions $W_1(\tau)$ and $W_2(\tau)$ it holds that
\begin{align}
\frac{\det\left[-\frac{d^2}{d\tau^2}+W_1(\tau)\right]}{\det\left[-\frac{d^2}{d\tau^2}+W_2(\tau)\right]}=\frac{\psi^{(1)}_0(T/2)}{\psi^{(2)}_0(T/2)}.
\end{align}
Let us take $W_1(\tau)=1$ and $W_2(\tau)=V''(g^{-1/2}\bar x(\tau))$.
One can check that the solution of (\ref{eq:Liouville_eq}) for $i=1, \lambda=0$ is $\psi^{(1)}_0(\tau)=\sinh(\tau+T/2)$. The two solutions of the equation $L_2 y(\tau)=0$ are
\begin{align}\begin{split}
y_1(\tau)&=\cosh^{-1}(\tau),\\
y_2(\tau)&=\sinh(\tau)+\tau\cosh^{-1}(\tau).
\end{split}\end{align}
Function $\psi^{(2)}_0(\tau)$ is a superposition of those and for large $T$ it reads $\psi^{(2)}_0(\tau)\approx\frac{1}{4}e^{T/2} y_1(\tau)+e^{-T/2} y_2(\tau)$.
Green's function for operator $L_2$ is
\begin{align}
G(\tau,\tau')&=
\left\{\begin{array}{ll}
-\mathcal W^{-1}y_1(\tau')y_2(\tau)&\text{for } \tau>\tau'\\
-\mathcal W^{-1}y_1(\tau)y_2(\tau')&\text{for } \tau<\tau'
\end{array}
\right.
\end{align}
where $\mathcal W\equiv y_1(\tau)\dot y_2(\tau)-\dot y_1(\tau)y_2(\tau)=-2$ is the Wronskian.
Let $\lambda_0$ be the smallest eigenvalue $L_2$. Then,
\begin{align}
0&=\psi^{(2)}_{\lambda_0}(T/2)=\psi^{(2)}_0(T/2)-\int_{-T/2}^{T/2}d\tau'G(T/2,\tau')\lambda_0\psi^{(2)}_{\lambda_0}(\tau')\approx1-\frac{e^T}{8}\lambda_0,
\end{align}
where we used the fact that $\psi^{(2)}_{\lambda_0}\approx\psi^{(2)}_0$ to calculate the integral. It follows that $\lambda_0\approx 8e^{-T}$.
Finally,
\begin{align}
\mathcal K=\sqrt{\lambda_0}\left(\frac{\det\left[-\frac{d^2}{d\tau^2}+1\right]}{\det\left[-\frac{d^2}{d\tau^2}+V''(g^{-1/2}\bar x(\tau))\right]}\right)^{1/2}=\sqrt{\lambda_0}\left(\frac{\psi^{(1)}_0(T/2)}{\psi^{(2)}_0(T/2)}\right)^{1/2}=2.
\end{align}

\end{document}